\documentclass[twocolumn,showpacs,preprintnumbers,amsmath,amssymb]{revtex4}

\usepackage{graphicx}
\usepackage{dcolumn}
\usepackage{bm}
\usepackage{xspace} 

\newcommand{\ket}[1]{|{#1}\rangle}

\def\trace#1{\text{Tr}({#1})}
\def\braket#1{\mathinner{\langle{#1}\rangle}}

\def\Braket#1#2#3{\langle {#1} | {#2} |{#3}\rangle}

\def\real{\text{Re}}
\def\imag{\text{Im}}
\def\Rb87{$^{87}\text{Rb}$}
\def\0{\ket{0}}
\def\1{\ket{1}}

\begin{document}

\preprint{APS/123-QED}

\title{The distribution of quantum fidelities}

\author{Line Hjortsh\o j Pedersen$^1$, Niels Martin M\o ller$^2$ and Klaus M\o lmer$^1$}
\affiliation{$^1$Lundbeck Foundation Theoretical Center for Quantum System Research, Department of
Physics and Astronomy, University of Aarhus, DK-8000 \AA rhus C, Denmark.\\
$^2$Department of Mathematical Sciences, University of Aarhus,
Ny Munkegade, Bld. 1530, DK-8000 \AA rhus C, Denmark}

\date{\today}

\begin{abstract}
When applied to different input states, an imperfect quantum operation yields output states with varying fidelities, defined as the absolute square of their overlap with the desired states. We present an expression for the  distribution of fidelities for a class of operations applied to a general qubit state, and we present general expressions for the variance and input-space averaged fidelities of arbitrary linear maps on finite dimensional Hilbert spaces.
\end{abstract}

\pacs{03.67.-a, 42.50.Lc, 89.70.+c}
\maketitle

\section{Introduction}

In quantum control and quantum information theory one attempts to
control the dynamics of a quantum system, such that the net
mapping of a quantum state by the dynamics yields a specific output
state. If the system is known to initially populate a specific
state, the success of a given operation on that state is determined
by the square of the overlap between the final state and the desired
output state, while, more generally, the successful implementation
of a certain map, e.g., a quantum gate in a quantum computer, should
be judged by an evaluation of this overlap for all possible input
states. Imperfections may be due to a variety of reasons, such as
dissipative coupling to auxiliary degrees of freedom and imprecise
knowledge of the physical parameters characterizing the system. A
wide range of methods have thus been applied to counter such
effects: composite pulses, quantum control, bang-bang control,
error-correcting codes, and use of decoherence-free subspaces.T
o optimize these methods it is necessary to have definite
functionals of the gate operation, that one can determine and,
hopefully, improve by suitable variation of accessible control
parameters.

In this paper, we address the distribution of fidelities obtained
when the input states are taken uniformly from the full  Hilbert
space of the physical system or from a suitable subspace, to which
physical circumstances may restrict the initial state. In Sec. II we
present a derivation of the fidelity distribution of unitary gates
applied to a qubit, i.e., a two-level quantum system, and an
extension to the case of a unitarily diagonalizable linear map. In
Sec. III, we review  our recent derivation \cite{PMM07} of the mean
value of the fidelity of an arbitrary linear map, and in Sec. IV, we
determine the variance of the fidelity distribution of such linear
maps. Sec. V concludes the paper.

\section{Fidelity distribution for a one-qubit gate}

Consider a two-level system, a qubit, subjected to a unitary
operator $U$, while the desired operation on the system is given by
the unitary operator $U_0$. Under ideal circumstances, the unitary
operator $M=U_0^\dagger U=U_0^{-1}U$ is the identity operator, and
the fidelity of the quantum operation $|\langle\psi|U_0^\dagger
U|\psi\rangle|^2=1$ for all normalized states $|\psi\rangle$. Under
less ideal circumstances, however, $M$ is not the identity, but it
can be diagonalized by a unitary transformation, and has two complex
eigenvalues $e^{i\phi_0}$ and $e^{i\phi_1}$, with $\phi_0 \neq
\phi_1$. We observe that the fidelity of the gate operation is unity
when $\psi$ is equal to either of the eigenvectors $|0\rangle,\
|1\rangle$ of $M$, while expanding $|\psi\rangle$, we get the lowest
fidelity,
$f=|(e^{i\phi_0}+e^{i\phi_1})/2|^2=\cos^2\left((\phi_1-\phi_0)/2\right)$,
for any equal weight superposition of these eigenvectors. In a Bloch
sphere picture with the eigenvectors as the north and south pole,
the fidelity is only a function of the polar angle $\theta$ of the
input state, $|\psi\rangle = \cos(\theta/2)|0\rangle +e^{i\phi}
\sin(\theta/2) |1\rangle$,
$f(\theta)=|e^{i\phi_0}\cos^2(\theta/2)+e^{i\phi_1}\sin^2(\theta/2)|^2$,
and with a uniform distribution of  states over the surface of the
Bloch sphere, and the corresponding polar angle distribution
$P_\theta = \frac{1}{2}\sin\theta$ one readily determines the
fidelity distribution, $P_f=\sum_i |df(\theta)/d\theta|_i^{-1}
P_{\theta_i}$, where the contributions to the sum come from
symmetric polar angles above and below the equator with the same
fidelity.

The expression for the fidelity distribution of a unitary, erroneous qubit gate thus reads,
\begin{equation} \label{funit}
P_f = \frac{1}{2\sin\left( \frac{\phi_1-\phi_0}{2}\right)\sqrt{f-\cos^2\left( \frac{\phi_1-\phi_0}{2}\right)}},
\end{equation}
for $\cos^2\left((\phi_1-\phi_0)/2\right) \leq f \leq 1$.

We shall not attempt a derivation of the equivalent distribution of
fidelities in higher Hilbert space dimensions. A related problem,
dealing with the distribution of matrix elements of a Hermitian
operator, was studied by von Neumann \cite{N41}, and illustrates how
the solution of such a problem breaks up in a large number of
different cases. We shall, however, extend our analysis to the
special situation where a quantum system is known to initially populate a
two-dimensional subspace of a three-dimensional Hilbert space,
including an excited state with suitable interaction properties,
such that excitation from the lower qubit states to the excited
states can be used to communicate between different quantum systems,
e.g., by an excited state dipole-dipole interaction. In \cite{RM04}
a robust one-qubit gate scheme was proposed, in which both qubit
states are simultaneously coupled to the excited state, giving rise
to a dark state superposition, $\ket{\bar{0}}$, and a bright state
superposition, $\ket{\bar{1}}$, with destructive and constructive
interference of the couplings. Transferring the bright state into
the excited state and back with laser fields with different phases
implements a controllable phase in the dark/bright basis, equivalent
to an arbitrary qubit rotation in the computational qubit basis
\cite{RM04}. If the coherent coupling of the bright and excited
state is imprecise and does not return the population fully to the
low lying states, we obtain $U = \Bigl( \begin{smallmatrix} 1 & 0 &
0 \\ 0 & \alpha & \gamma \\ 0 & \gamma* & \beta \end{smallmatrix}
\Bigr)$ in the basis $\{\ket{\bar{0}},\ket{\bar{1}},\ket{e}\}$.
This implies that the effect on the qubit space of the application of $U$
can be written in terms of $U$ and the projection operator $P$ on that space,
$PUP = \bigl( \begin{smallmatrix} 1 & 0 \\ 0 &
\alpha \end{smallmatrix} \bigr)$ in the basis
$\{\ket{\bar{0}},\ket{\bar{1}}\}$. $\alpha$ is a complex number, and
if $|\alpha|= 1$ the fidelity distribution is given by Eq. (\ref{funit})
with the difference between $\arg(\alpha)$ and the desired phase shift by $U_0$
replacing $\phi_1-\phi_0$.

 An operation with $|\alpha|< 1$ yields a special example of the slightly more general case, where the state vector is mapped, $|\psi\rangle \mapsto N|\psi\rangle$, by a normal 2 x 2 matrix $N$. The matrix $M=U_0^\dagger N$ is also normal, i.e., it can be diagonalized by a unitary transform with eigenvalues $\lambda_0, \lambda_1$ that we can arrange such that $|\lambda_0|\leq |\lambda_1|$. Employing the expansion on the corresponding eigenstates, parametrized by
$c_0 = \cos(\theta/2)$ and $c_1 = e^{i\phi}\sin(\theta/2)$, where $\phi \in [0,2\pi]$ and
$\theta \in [0,\pi]$, we have $f = |\lambda_0|^2 f_\lambda$, where
\begin{equation}\label{flambda}
f_\lambda = \cos^4 \frac{\theta}{2} + \sin^4\frac{\theta}{2}|\lambda|^2 + \cos^2 \frac{\theta}{2} \sin^2\frac{\theta}{2}(\lambda+\lambda^*),
\end{equation}
and $\lambda = \lambda_1/\lambda_0$. With the same argument as above we need
\begin{equation}\label{eq:dfdtheta}
\frac{df}{d\theta} = \frac{1}{2}|\lambda_0|^2\sin\theta (|\lambda|^2-1-|1-\lambda|^2\cos\theta),\nonumber
\end{equation}
and we obtain
\begin{equation}\label{eq:pftheta}
|\lambda_0|^2 P_{f} = \left| |\lambda|^2-1-|1-\lambda|^2\cos\theta\right|^{-1},\nonumber
\end{equation}
where, from \eqref{flambda},
\begin{equation}\label{costheta}
\cos\theta = \frac{|\lambda|^2-1\pm 2\sqrt{f_\lambda|1-\lambda|^2-\imag(\lambda)^2}}{|1-\lambda|^2}.\nonumber
\end{equation}
The fidelity is a non-monotonic function of the polar angle
$\theta$, and adding the contributions from different polar angles
yielding the same fidelity, we finally obtain
the probability distribution (recall $|\lambda_0|^2 < |\lambda_1|^2$).
We have to specify the results for two different cases.\\

\noindent
If $|\lambda_0-\frac{1}{2}\lambda_1|<\frac{1}{2}|\lambda_1|$, then
\begin{equation}\label{distnocrit}
P_f  = \frac{1}{2|\lambda_0-\lambda_1|}\frac{1}{\sqrt{f-f_0}}, \qquad |\lambda_0|^2 \leq f \leq |\lambda_1|^2,
\end{equation}

\noindent
and if  $|\lambda_0-\frac{1}{2}\lambda_1|\geq \frac{1}{2}|\lambda_1|$, then
\begin{equation}\label{distcrit}
P_f =
\begin{cases}
\frac{1}{|\lambda_0-\lambda_1|}\frac{1}{\sqrt{f-f_0}}, & \quad f_0\leq f \leq |\lambda_0|^2, \\
\frac{1}{2|\lambda_0-\lambda_1|}\frac{1}{\sqrt{f-f_0}}, & \quad |\lambda_0|^2 \leq f \leq |\lambda_1|^2,
\end{cases}
\end{equation}
where $f_0 = \imag(\lambda_0\lambda_1^*)^2/|\lambda_0-\lambda_1|^2$.

Fig. 1. shows the fidelity distribution (\ref{distcrit}) for the
case of $\lambda_0= 0.7\cdot e^{i\pi/8},\ \lambda_1= 0.8\cdot
e^{i4\pi/5}$. The vertical bars in the figure indicate a histogram
obtained by drawing states on the Bloch sphere at random and binning
their individual fidelities. We observe that the results agree very
well.

\begin{figure}[t]
\centering {\includegraphics[width=7cm]{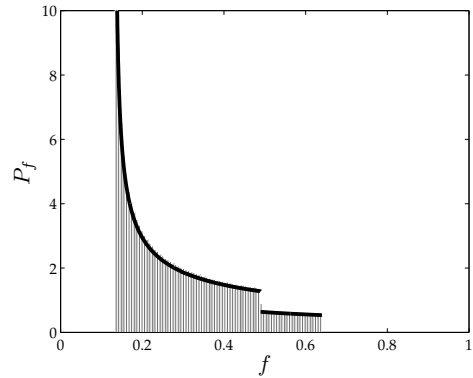}}
\caption{Theoretical (line) and numerical (histogram) fidelity
distribution for a matrix with eigenvalues $\lambda_0 = 0.7\cdot
e^{i\pi/8}$ and $\lambda_1 = 0.8\cdot e^{i4\pi/5}$. In accordance with \eqref{theorem}
the average fidelity is 0.28.
} \label{fig:fidelitydist}
\end{figure}

Having the full fidelity distribution at our disposal, we can see
how changes in the map, reflected in the parameters
$\{\lambda_{0},\lambda_1\}$, change the fidelities, and we can
optimize according to different criteria: minimum value, mean value
or some higher moment, which can readily be obtained from the
distribution.

In the following, we shall limit our analysis of the fidelity distribution to a characterization of its mean
value and variance. Rather than obtaining these numbers from
integrals over the distribution (\ref{distnocrit},\ref{distcrit}),
 $\braket{f} = \int_{f_{min}}^{f_{max}} P_f f df$ and
$\braket{f^2} = \int_{f_{min}}^{f_{max}} P_f f^2 df$, we shall provide a more general expression, valid for any map $|\psi\rangle \mapsto N |\psi\rangle$, in any finite dimensional setting.

\section{The average fidelity measure}\label{sec:averagefidelity}

Derivations have been given in the literature for the average fidelity of qubit \cite{HHH99}
and qudit operations \cite{N02,PMM07}, as well as for their evaluation as
a sum over a properly chosen discrete set of states
\cite{BOSBJ02,BBM03}. The latter approach is particularly relevant
in connection with quantum process tomography \cite{NC00}, which
provides a procedure for determining the quantum operation acting on a
system from experimental observations.\\

For any linear operator $M$ acting on an $n$-dimensional complex Hilbert
space, the uniform average of $|\langle \psi | M| \psi \rangle |^2$
over state vectors $|\psi\rangle$ on the unit sphere $S^{2n-1}$ in
$\mathbb{C}^n$ is given by
\begin{equation}\label{theorem}
\begin{split}
& \langle f \rangle = \int_{S^{2n-1}} |\langle \psi | M| \psi \rangle |^2 dV \\
& = \frac{1}{n(n+1)}\big[\text{Tr}(MM^\dagger)+|\text{Tr}(M)|^2\big],
\ \
\end{split}
\end{equation}
where $dV$ is the normalized measure on the sphere.

Recent detailed proofs of this result are given in \cite{ZL04,D05,PMM07},
and it is also readily verified using a recent result for the
averages of general polynomials of state vector amplitudes over the
unit sphere \cite{AE07}.

The expression (\ref{theorem}) is readily evaluated for any matrix
$M$, and in particular, for the qubit example with eigenvalues
$\lambda_{0}$ and $\lambda_1$, we get the simple result, $\braket{f} =
\left(|\lambda_0|^2+|\lambda_1|^2+\real
(\lambda_0\lambda_1^*)\right)/3$. If $M$ is unitary, $MM^\dagger$ is
the identity matrix with trace equal to the dimension $n$ in Eq.(5).

In the following we recall a few useful results, derived in \cite{PMM07}.

\subsection{Subspace averaged fidelity of a unitary transformation}\label{sec:subavfid}

In a number of quantum information scenarios, auxiliary quantum
levels are used to mediate the desired operations. In these
protocols, the auxiliary levels of the quantum system may, with
certainty, be unpopulated before the process and, consequently, we
should only average the fidelity over the relevant input states.
Since the final state is, ideally, also in the same subspace, we
consider the matrix $M=(PU_0^\dagger P)(PUP)$, where $P$ is the
projection operator on the relevant, quantum information carrying
subspace $\mathbb{S}$, and $U_0$ represents the desired unitary
evolution of the relevant subspace. In a matrix notation, the
outermost applications of the projection operator $P$ amounts to the
extraction of the square $n_{rel} \times n_{rel}$ submatrix
$M_{rel}$ with the relevant columns and rows of the full matrix $M$,
and to compute the mean fidelity over the subspace, we employ
(\ref{theorem}) for this reduced matrix:
\begin{equation}\label{fidrest}
\begin{split}
F & =\frac{1}{n_{rel}(n_{rel}+1)}\big[\text{Tr}(M_{rel}M_{rel}^\dagger)+|\text{Tr}(M_{rel})|^2\big],
\end{split}
\end{equation}
If population leaks to the auxiliary levels, $M_{rel}$ is
not unitary, and hence both terms of \eqref{fidrest} have nontrivial
values.

If a measurement assures that the final state does not populate the
complement to $\mathbb{S}$, it is meaningful to define the average
\emph{conditional} fidelity $F_c$. If we only accept the state if it is
in $\mathbb{S}$, the squared overlap between the conditional,
renormalized state $\frac{PUP\ket{\psi}}{\| PUP\ket{\psi}\|}$ and
the ideal state $U_0 P\ket{\psi}$ must be weighted with the acceptance probability
$\| PUP\ket{\psi}\|^2$ and renormalized by the integrated acceptance probability over the input Hilbert space, $\int \| PUP\ket{\psi}\|^2 dV$. We hence obtain
the average conditional fidelity \footnote{This expression replaces
an incorrect expression in \cite{PMM07}.}
\begin{equation*}
\begin{split}
F_c & = \int |\Braket{\psi}{PU_0^\dagger \frac{P U P}{\| PUP\ket{\psi}\|}}{\psi}|^2\cdot \frac{\| PUP\ket{\psi}\|^2 }{\int \| PUP\ket{\psi}\|^2 dV} dV\\
& = \frac{\int |\Braket{\psi}{PU_0^\dagger PUP}{\psi}|^2 dV}{\int \Braket{\psi}{PU^\dagger PUP}{\psi} dV} \\
& = \frac{1}{n_{rel}+1}\cdot\frac{\trace{U_0^\dagger PUPU^\dagger PU_0} + |\trace{U_0^\dagger PUP}|^2}{\trace{U^\dagger PUP}},
\end{split}
\end{equation*}
where the numerator is evaluated using (\ref{theorem}) and the denominator follows from
$\int \Braket{\psi}{M}{\psi} dV = \trace{M}/n$.

\subsection{Average fidelity of a general quantum operation}

Our quantum system may not be fully isolated from its surroundings, and ancillary quantum systems may play significant roles in various quantum information processing scenarios: quantum memory
protocols in a very explicit manner involve an extra quantum system,
quantum teleportation requires an extra entangled pair of systems,
and in quantum computing with trapped ions, a motional degree of
freedom is used to couple the particles. The ancillary systems are
ideally disentangled from the qubits before and after the process,
but in general they act as an environment and cause decoherence of
the quantum system of interest. This forces us to generalize the
formalism and take into account the general theory of
quantum operations, according to which the mean dynamics is accounted for by density matrices which
transform by completely positive maps. According to the Kraus
representation theorem, any completely positive trace-preserving map
$\mathcal{G}$ admits the representation
\begin{equation}\label{eq:krausform}
\mathcal{G}(\rho) = \sum_k G_k \rho G_k^\dagger,
\end{equation}
where $\sum_k G_k^\dagger G_k = I_n$ is the $n \times n$ identity
matrix \cite{NC00}.

If the pure input state $\rho = |\psi\rangle \langle \psi|$ is
mapped to the output state $\mathcal{G}(\rho)$ the mean fidelity
with which our operation yields a unitary transformation $U_0$ is
\begin{equation}\label{cor}
\begin{split}
F &= \int_{S^{2n-1}} \langle \psi | U_0^\dagger \mathcal{G}(|\psi\rangle\langle\psi|) U_0|\psi\rangle dV \\
&= \sum_k \int_{S^{2n-1}} |\langle \psi
|M_k|\psi\rangle|^2 dV\\
& = \frac{1}{n(n+1)} \Bigl\{ \text{Tr} \Bigl(\sum_k M_k^\dagger M_k
\Bigr)+\sum_k |\text{Tr} (M_k)|^2 \Bigr\},
\end{split}
\end{equation}
where $M_k = U_0^\dagger G_k$.

Equation \eqref{cor} enables the calculation of the average fidelity
of any quantum operation, as soon as it has been put in the Kraus
form. For examples, see \cite{PMM07}.

\section{The variance of the fidelity distribution}\label{sec:varfid}

To further characterize the fidelity distribution, we shall now
obtain an explicit formula for the variance of the fidelity
distribution. Note that it is not meaningful to define the variance
of a general quantum transformation governed by the Kraus form,
since the density matrix formulation already incorporates an averaging
procedure, due to the trace over unobserved degrees of freedom of
the surroundings of the quantum system. The density matrix evolution
may be understood, and simulated, by an average over randomly
evolving wave functions \cite{MCWF}, but this unravelling is not
unique \cite{MCWF2}, and only if measurements are actually carried out on the
surroundings, a specific stochastic dynamics of the state
vectors happens, and it is possible to assign a definite fidelity distribution,
and variance of this distribution, to a given process.

We shall proceed and derive a relation for the uniform average
of $f^2$, i.e., $|\langle \psi | M| \psi \rangle |^4$, with $M=U_0^\dagger N$, for linear
state vector maps $|\psi\rangle \mapsto N|\psi\rangle$, where $N$ is
any linear operator.

First, we show that for a Hermitian operator $S$,
the uniform average of $|\langle \psi | S| \psi \rangle |^4$
over state vectors $|\psi\rangle$ on the unit sphere $S^{2n-1}$ in
$\mathbb{C}^n$ is given by
\begin{equation}\label{hervareq}
\begin{split}
& \int_{S^{2n-1}} |\langle \psi | S| \psi \rangle |^4 dV  =
\frac{1}{n(n+1)(n+2)(n+3)} \times\\
& \quad \big[6\trace{S^4}+8\trace{S^3}\trace{S}+3\trace{S^2}^2\\
& \quad +6\trace{S^2}\trace{S}^2+\trace{S}^4].
\end{split}
\end{equation}

Our demonstration of this result proceeds along the lines of our
demonstration of Eq.(\ref{theorem}), given in more detail in \cite{PMM07} . First we note the
invariance of both sides of \eqref{hervareq} under conjugation by any unitary operator $S \rightarrow U^\dagger SU$, which allows us restrict the analysis to a diagonal
matrix $\Lambda$ with real eigenvalues $\lambda_1,\ldots,\lambda_n$. Letting $L$ denote the left-hand side of
\eqref{hervareq}, we observe that $L(\Lambda)$ is a homogeneous
polynomial of degree 4 in the real variables
$\lambda_1,\ldots,\lambda_n$, which is invariant under the exchange
of any two $\lambda_i$ and $\lambda_j$. It is easy to demonstrate
that the set
\begin{equation*}
\left\{\trace{\Lambda^4},\trace{\Lambda^3}\trace{\Lambda},
\trace{\Lambda^2}^2,\trace{\Lambda^2}\trace{\Lambda}^2,\trace{\Lambda}^4\right\}
\end{equation*}
spans all such polynomials $L(\Lambda)$, and by  evaluating the integrals
\begin{equation}\label{integrals}
\begin{split}
&\int_{S^{2n-1}} |c_i|^8 dV = \frac{24}{n(n+1)(n+2)(n+3)} \\
&\int_{S^{2n-1}} |c_i|^4 |c_j|^4 dV = \frac{4}{n(n+1)(n+2)(n+3)} \\
&\int_{S^{2n-1}} |c_i|^6 |c_j|^2 dV = \frac{6}{n(n+1)(n+2)(n+3)} \\
&\int_{S^{2n-1}} |c_i|^4 |c_j|^2 |c_k|^2 dV = \frac{2}{n(n+1)(n+2)(n+3)} \\
&\int_{S^{2n-1}} |c_i|^2 |c_j|^2 |c_k|^2 |c_l|^2 dV =
\frac{1}{n(n+1)(n+2)(n+3)},
\end{split}
\end{equation}
for different $i,j,k,l$, and by choosing choosing five
appropriate matrices for which the integrals are readily obtained, we finally obtain the coefficients in
\eqref{hervareq} by
solving a linear system of equations.

Note that \eqref{hervareq} also holds for an
anti-Hermitian matrix $A$, since $L(A) = L(iA) = R(iA) = R(A)$,
where $L$ and $R$ denote the left- and right-hand sides of
\eqref{hervareq}, respectively.

Having obtained the uniform average of $|\langle \psi | S| \psi
\rangle |^4$, where $S$ is Hermitian, we now proceed to the general case.
We decompose the general matrix $M=S+A$ as a sum of a Hermitian and an
anti-Hermitian matrix, denoted by $S$ and $A$, respectively, and we note
that
\begin{multline}\label{sadecomp}
\int_{S^{2n-1}} |\Braket{\psi}{M}{\psi}|^4 dV = \\ \int_{S^{2n-1}}
|\Braket{\psi}{S}{\psi}|^4+ |\Braket{\psi}{A}{\psi}|^4 +
2|\Braket{\psi}{S}{\psi}|^2 |\Braket{\psi}{A}{\psi}|^2 dV.
\end{multline}
The first two terms on the right-hand side of \eqref{sadecomp} are
readily evaluated using (\ref{hervareq}) and the explicit
expressions $S = (M+M^\dagger)/2$ and $A = (M-M^\dagger)/2$.

To calculate the third term, we use the conjugation invariance to diagonalize the Hermitian part $S$, and thus
\begin{equation}\label{latilde}
\begin{split}
& \int_{S^{2n-1}} |\Braket{\psi}{S}{\psi}|^2
|\Braket{\psi}{A}{\psi}|^2 dV = \\ & \qquad \int_{S^{2n-1}}
|\Braket{\psi}{\Lambda}{\psi}|^2 |\Braket{\psi}{\tilde{A}}{\psi}|^2
dV,
\end{split}
\end{equation}
where $\Lambda = USU^{-1}$ is a diagonal matrix with elements
$\lambda_1,\ldots,\lambda_n \in \mathbb{R}$, and $\tilde{A} =
UAU^{-1}$ is an anti-Hermitian matrix, which is not necessarily
diagonal. Expanding the state vectors in the eigenbasis of $S$,
$\ket{\psi} = \sum c_j \ket{j}$ and employing the invariance of the
integral \eqref{latilde} under the transformation $c_j \rightarrow
\exp(i\theta_j) c_j$, it follows that
\begin{equation}\label{saterms}
\begin{split}
& \int_{S^{2n-1}} |\Braket{\psi}{\Lambda}{\psi}|^2
|\Braket{\psi}{\tilde{A}}{\psi}|^2 dV = \\ & \; \; \int_{S^{2n-1}}
\sum_{\alpha,\beta,i,j,k,l} |c_\alpha|^2 |c_\beta|^2 c_i^* c_l^{*}
c_j c_k \lambda_\alpha \lambda_\beta \tilde{A}_{ij} \tilde{A}_{kl}^*
\times \\ & \; \; (\delta_{ij}\delta_{kl} + \delta_{ik} \delta_{jl}
- \delta_{ij}\delta_{kl}\delta_{il}) dV = \frac{1}{n(n+1)(n+2)(n+3)}
\times \\ & \; \; \Big\{ 24\sum_i \lambda_i^2 |\tilde{A_{ii}}|^2 +
\sideset{}{'} \sum_{ijkl} \lambda_k\lambda_l
(\tilde{A}_{ii}\tilde{A}_{jj}^*+|\tilde{A}_{ij}|^2)+\\ & \; \;
4\sideset{}{'}\sum_{ij} \left[ (\lambda_j^2 +
3\lambda_i\lambda_j)|\tilde{A}_{ii}|^2 +
(3\lambda_i^2+2\lambda_i\lambda_j)(\tilde{A}_{ii}\tilde{A}_{jj}^*+|\tilde{A}_{ij}|^2)\right]
\\ & \; \; +2\sideset{}{'}\sum_{ijk} \left[
\lambda_i\lambda_j |\tilde{A}_{kk}|^2 +
(\lambda_k^2+4\lambda_i\lambda_k)(\tilde{A}_{ii}\tilde{A}_{jj}^*+|\tilde{A}_{ij}|^2)\right]
\Big\},
\end{split}
\end{equation}
where the last step follows upon insertion of the expressions obtained in \eqref{integrals}. The notation $\sideset{}{'} \sum$
indicates that all indices must be different.
After a lengthy, but straightforward calculation, \eqref{saterms} can be rewritten in terms of traces of products of powers of $\Lambda$ and $\tilde{A}$. Invoking trace invariance, we can
replace $\Lambda$ and $\tilde{A}$ with $S$ and $A$, respectively, and insert the explicit expressions
for $S$ and $A$ in terms of $M$ and $M^\dagger$. Collecting terms in \eqref{sadecomp}, we finally obtain the man value of the squared fidelity, i.e., the uniform
average of $|\langle \psi | M| \psi \rangle |^4$ for the
linear operator $M$,
\begin{equation}\label{theoremvareq}
\begin{split}
\int_{S^{2n-1}} & |\langle \psi | M| \psi \rangle |^4 dV  =
\frac{1}{n(n+1)(n+2)(n+3)}\times \\
& \big[4\trace{M^2M^{\dagger2}}+2\trace{MM^\dagger MM^\dagger} \\
& +4\trace{M}\trace{MM^{\dagger 2}}+4\trace{M^{\dagger}}\trace{M^2M^\dagger}\\
& +\trace{M^2}\trace{M^{\dagger2}}+2\trace{MM^\dagger}^2\\
& +\trace{M^2}\trace{M^\dagger}^2+\trace{M}^2\trace{M^{\dagger2}} \\
& +4\trace{M}\trace{M^{\dagger}}\trace{MM^{\dagger}}+\trace{M}^2\trace{M^{\dagger}}^2
\big].
\end{split}
\end{equation}

\noindent Combining (\ref{theorem}) and (\ref{theoremvareq}), we
obtain the variance of the fidelity distribution for a linear
transformation $|\psi\rangle \mapsto N|\psi\rangle$, using
\begin{equation*}
\begin{split}
\sigma_f^2 & = \langle f^2\rangle - \langle f^2\rangle\\ & =
\int_{S^{2n-1}} |\langle \psi | M| \psi \rangle |^4 dV - \left(
\int_{S^{2n-1}} |\langle \psi | M| \psi \rangle |^2 dV \right)^2,
\end{split}
\end{equation*}
with $M = U_0^\dagger N$, where $U_0$ and $N$ are the desired and
actual evolution operators of the system. As in Sec.
\ref{sec:subavfid}, this result is easily extended to the case where
the average is solely performed over a subset of input states by
simply replacing $n$ with $n_{rel}$, and $M$ with $M_{rel} =
PU_0^\dagger PNP$, where $P$ is the corresponding projection operator
on the relevant subspace.

\section{Summary}

We have in this paper derived an expression for the fidelity
distribution, applicable to a normal linear state vector
transformation in a two dimensional complex Hilbert space. We have
also presented simple and compact expressions for the average
fidelity of a general quantum operation, and for the variance of
the fidelity of a linear state vector map. Such
simple expressions constitute a good starting point for further analysis,
e.g., for the achievements of error correcting codes \cite{S99},
decoherence free subspaces \cite{ZR97,LCW98}, and protection of
quantum information by dynamical decoupling \cite{VKL99}. Our
expression can also be handled and generalized analytically, as
illustrated by our study in \cite{PMM07} of a $K$-qudit register,
which provides insight into the scaling of errors. This may have
applications in quantum error correction, the capacity of quantum
channels, and the way that, e.g., communication with quantum
repeaters \cite{DLCZ01} and entanglement distillation should
optimally be carried out. Although we did not consider that
possibility here, we note that the ability to restrict averages to
subspaces may also enable generalization of our formalism to deal
with non-uniform averages, assuming nontrivial prior probability
distributions on the Hilbert space.

\end{document}